\documentclass{article}
\usepackage[a4paper, left=25mm, right=25mm, top=25mm, bottom=20mm, footskip=30pt, headsep=25pt]{geometry} 
\usepackage{amsmath,amsfonts}
\usepackage{array}
\usepackage[caption=false,font=normalsize,labelfont=sf,textfont=sf]{subfig}
\usepackage{textcomp}
\usepackage{stfloats}
\usepackage{url}
\usepackage{verbatim}
\usepackage{graphicx}
\usepackage{cite}

\usepackage{stmaryrd} 
\usepackage{bm} 
\usepackage{amssymb} 
\usepackage[ruled, vlined, linesnumbered]{algorithm2e}
\usepackage{placeins} 
\usepackage{accents} 
\usepackage[svgnames]{xcolor} 
\usepackage{tikz} 
\usepackage{tikz-3dplot} 
\usepackage{pgfplots} 
\usetikzlibrary{arrows.meta,positioning,shapes.geometric,shapes.misc,fit,calc} 
\pgfplotsset{compat=1.18} 
\usepackage{enumitem} 
\usepackage{hyperref} 
\usepackage{empheq} 

\newcommand{\ent}[2]{\llbracket #1,#2 \rrbracket}

\newcommand{\A}{{\mathbf{A}}} 
\newcommand{\R}{{\mathbb{R}}}

\newcommand{\x}{{\bm{x}}}

\newcommand{\y}{{\bm{y}}}

\newcommand{\norm}[2]{{\| #2 \|}_{#1}}

\newcommand{\bfone}{\bm{1}}
\newcommand{\bfzero}{\bm{0}}
\newcommand{\lbf}{\bm{\ell}}
\newcommand{\Hmat}{{\mathbf{H}}} 
\newcommand{\h}{{\bm{h}}}

\usepackage{amsthm} 
\theoremstyle{plain} 

\theoremstyle{definition} 

\theoremstyle{remark} 

\usepackage{tcolorbox} 
\tcbuselibrary{breakable, skins, vignette}

\DeclareMathOperator*{\argmin}{arg \, min}

\usepackage{authblk}

\title{An active-set algorithm for spectral unmixing}
\author[1]{Nils Foix-Colonier}
\author[1]{Sébastien Bourguignon}
\affil[1]{Nantes Université, École Centrale Nantes, CNRS, LS2N, UMR 6004, F-44000 Nantes, France}
\date{}

\begin{document}
\maketitle

\begin{abstract}
    Linear spectral unmixing under nonnegativity and sum-to-one constraints is a convex optimization problem for which many algorithms were proposed. In practice, especially for supervised unmixing (i.e., with a large dictionary), solutions tend to be sparse due to the nonnegativity of the abundances, thereby motivating the use of an active-set solver. Given the problem specific features, it seems advantageous to design a dedicated algorithm in order to gain computational performance compared to generic solvers. In this paper, we propose to derive such a specific algorithm, while extending the nonnegativity constraints to broader minimum abundance constraints.
\end{abstract}

\section{Problem definition}

Given a spectral library $\A\in\R^{N\times P}$, a measured spectrum $\y\in\R^N$, and a lower-bound vector $\lbf\in\R^P_+$ that is feasible, i.e. $\bfone^\mathsf{T} \lbf \le 1$, we consider the following unmixing problem:
\begin{equation}\label{eq:orig}
\min_{\x\in\R^P}\ \tfrac12\norm{2}{\y-\A\x}^2
\quad\text{s.t.}\quad
\x\ge \lbf,\;\; \bfone^\mathsf{T} \x=1
\end{equation}

For convenience, the variable is shifted: $\tilde \x = \x-\lbf$. We also define $\tilde \y = \y - \A\lbf$ and $s=1-\bfone^\mathsf{T}\lbf$. Then~\eqref{eq:orig} is  equivalent to:
\begin{equation}\label{eq:shifted}
\min_{\tilde \x \in \R^P} \;\; \tfrac12\norm{2}{\tilde \y-\A \tilde\x}^2
\quad\text{s.t.}\quad
\tilde \x \ge =\bfzero,\;\; \bfone^\mathsf{T} \tilde \x=s
\end{equation}

Consequently, solving~\eqref{eq:shifted} yields the solution to~\eqref{eq:orig}, by shifting back: $\x=\tilde \x+\lbf$. Note that, with $\lbf=\bfzero$, we have the standard Fully Constrained Least Squares (FCLS) formulation~\cite{fcls2001}.

\section{KKT conditions}

The objective function in~\eqref{eq:shifted} can be rewritten as the convex quadratic objective function:
\begin{equation}
f(\tilde{\x}):=\tfrac12 \tilde{\x}^\mathsf{T} \Hmat \tilde{\x} - \h^\mathsf{T} \tilde{\x} + \tfrac12\norm{2}{\tilde \y}^2 \text{, \; with $\Hmat:=\A^\mathsf{T} \A$ and $\h:=\A^\mathsf{T} \tilde{\y}$}.
\end{equation}

Let $\lambda\in\R$ be the Lagrange multiplier for the equality constraint $\bfone^\mathsf{T}\tilde{\x}=s$, and $\bm{\mu}\in\R^P$ the vector of multipliers for the non-negativity constraints $\tilde{\x}\ge \bfzero$. The Lagrangian function for problem~\eqref{eq:shifted} is:
\begin{equation}
    \mathcal{L}(\tilde{\x},\lambda,\bm{\mu})=f(\tilde{\x}) + \lambda(\bfone^\mathsf{T} \tilde{\x}-s) - \bm{\mu}^\mathsf{T} \tilde{\x}
\end{equation}
whose gradient reads:
\begin{equation}
    \nabla_{\tilde{\x}}\mathcal{L}(\tilde{\x},\lambda,\bm{\mu}) = \Hmat\tilde{\x} - \h + \lambda \bfone - \bm{\mu}.
\end{equation}

The Karush–Kuhn–Tucker (KKT) first-order conditions~\cite{nocedal2006} for  problem~\eqref{eq:shifted} read:
\begin{subequations}\label{eq:KKT}
\begin{empheq}[left=\empheqlbrace]{alignat=2}
&\Hmat\tilde{\x} - \h + \lambda \bfone - \bm{\mu} = \bfzero &\qquad& \text{\emph{(stationarity)}} \label{eq:KKT-sta}\\
&\bfone^\mathsf{T} \tilde{\x} = s,\ \tilde{\x}\ge \bfzero &\qquad& \text{\emph{(primal feasibility)}} \label{eq:KKT-pri}\\
&\bm{\mu} \ge \bfzero &\qquad& \text{\emph{(dual feasibility)}} \label{eq:KKT-dual}\\
&\forall i \in \ent{1}{P},\ \mu_i \tilde{x}_i = 0 &\qquad& \text{\emph{(complementary slackness)}} \label{eq:KKT-comp}
\end{empheq}
\end{subequations}

Since $f$ is convex and all constraints are convex, the first-order KKT conditions are a necessary and sufficient characterization of global optimality.

\section{Active-set algorithm}\label{sec3}


In order to find good candidates $\tilde \x$ for the KKT conditions~\eqref{eq:KKT}, we design an active-set algorithm, inspired by~\cite{nocedal2006}, based on the one introduced in 1974 by Lawson and Hanson~\cite{lawson1974}. It operates by iteratively determining which inequality constraints will be active at the solution and temporarily treats them as equalities. In our case, at any iterate with a feasible candidate $\tilde \x$, the indices of its coefficients are partitioned into an \emph{active set} $L=\{i\, | \, \tilde{x}_i=0\}$, for the variables currently at the lower bound, and a \emph{free set} $F=\{i\, | \,  \tilde{x}_i>0\}$ for the variables currently allowed to move inside the feasible space.

Let $S\subseteq\ent{1}{P}$, the notation $\A_S$ refers to the matrix $\A$ restricted to the columns indexed by $S$. In addition, let $\Hmat_{FF} = \A_F^\mathsf{T} \A_F$, $\Hmat_{LF} = \A_L^\mathsf{T} \A_F$.
According to the complementary slackness condition~\eqref{eq:KKT-comp} we have $\forall i \in F,\, \mu_i=0$. In addition, with dual feasibility~\eqref{eq:KKT-dual}: $\forall i \in L,\, \mu_i\geq0$. Consequently, with condition~\eqref{eq:KKT-sta}:
\begin{subequations}\label{eq:KKT-sta-FL}
\begin{empheq}[left=\empheqlbrace]{alignat=2}
&\Hmat_{FF} \tilde{\x}_{F} - \h_F + \lambda \bfone = \bfzero &\qquad& (\bm{\mu}_F=\bfzero) \label{eq:KKT-sta-F}\\
&\bm{\mu}_L = \Hmat_{LF} \tilde{\x}_F - \h_L + \lambda \bfone &\qquad& (\bm{\mu}_L\geq\bfzero) \label{eq:KKT-sta-L}
\end{empheq}
\end{subequations}
These equations are exploited in the active-set algorithm, which can be summarized in three steps:
\begin{enumerate}[leftmargin=2em]
\item \textbf{Solve an equality-constrained subproblem}, obtained by fixing $\tilde x_i=0$ for $i\in L$ (inequality constraints ignored for now). It consists of solving~\eqref{eq:KKT-sta-F} while enforcing $\bfone^\mathsf{T} \tilde \x=s$, which returns a candidate $\tilde \x_F^*$ and $\lambda^\star$. If  $\tilde \x_F^* \geq \bfzero$, then directly go to step 3. Otherwise, go to step 2.
\item \textbf{Keeping primal feasibility}: move from $\tilde \x$ towards $\tilde \x^*$, taking the largest step keeping nonnegativity constraints. The index of a variable that first saturates nonnegativity is moved from $F$ to $L$, and the algorithm goes back to step 1 with an updated~$\tilde \x$.
\item \textbf{Checking dual feasibility}:  compute the Lagrange multipliers $\bm{\mu}_L$ with equation~\eqref{eq:KKT-sta-L} evaluated with $\tilde \x^*$ and $\lambda^*$, from step 1. If $\bm{\mu}_L\geq \bfzero$, then $\tilde \x^*$ is a global minimizer and the algorithm stops. Otherwise, it fails to satisfy dual feasibility, so the index $i\in L$ corresponding to the most negative $\mu_i$ is moved from $L$ to $F$, and the algorithm moves back to step 1 with the updated candidate.
\end{enumerate}
This mechanism is directly motivated by the KKT conditions: the solve at step 1 enforces~\eqref{eq:KKT-sta} and the equality constraint from~\eqref{eq:KKT-pri}, while step 2 guarantees the primal feasibility on $F$ by respecting the inequalities of~\eqref{eq:KKT-pri} for $F$, and since the variables in the active set are null we have complete primal feasibility. The dual feasibility~\eqref{eq:KKT-dual} is verified at step 3, on $L$ only since the $\mu_i$ for $i\in F$ are null by construction. Finally, the complementary slackness~\eqref{eq:KKT-comp} is maintained by construction. We assume feasibility and nondegeneracy, ensuring unique minimizers for subproblems. Consequently, as the convergence under a finite number of iterations of such algorithms has been proven~\cite{nocedal2006}, we can guarantee that it will find a global minimizer of problem~\eqref{eq:shifted}.

\subsection{Solving an equality-constrained subproblem}\label{subsec1}

In order to solve equality-constrained subproblems (from step 1), we need to reformulate the problem at hand. We want to solve the equation system corresponding to the stationarity condition~\eqref{eq:KKT-sta}, under the sum equality constraint~\eqref{eq:KKT-pri}, and with respect to the constraints from the active set $L$. In practice, we do not need to solve the system in dimension $P$, since the coefficients with indices in $L$ are zero, so we just need to solve the reduced system~\eqref{eq:KKT-sta-F} under the sum equality constraint:

\begin{subequations}\label{eq:KKT-subpb}
\begin{empheq}[left=\empheqlbrace]{alignat=2}
&\Hmat_{FF} \tilde{\x}_F - \h_F + \lambda \bfone = \bfzero \label{eq:KKT-subpba}\\
& \bfone^\mathsf{T} \tilde{\x}_F = s\label{eq:KKT-subpbb}
\end{empheq}
\end{subequations}

This can be reformulated equivalently as:
\begin{equation}\label{eq:KKT-subpb-mat}
\begin{bmatrix}
\Hmat_{FF} & \bfone\\[2pt]
\bfone^\mathsf{T} & 0
\end{bmatrix}
\begin{bmatrix}
\tilde{\x}_F\\ \lambda
\end{bmatrix}
=
\begin{bmatrix}
\h_F\\ s
\end{bmatrix}
\end{equation}

Assuming that $\A_F$ has $|F|\leq N$ linearly independent columns, $\Hmat_{FF}$ is symmetric positive definite, therefore invertible, and its inverse is also symmetric positive definite. 
The determinant of the matrix in~\eqref{eq:KKT-subpb-mat} can be decomposed into~\cite{matcook}:
 \begin{equation}\label{eq:det}
 	\det \left(\begin{bmatrix}\Hmat_{FF} & \bfone\\[2pt]\bfone^\mathsf{T} & 0\end{bmatrix}\right) = \det \left(\Hmat_{FF} \right) \det \left(0- \bfone^\mathsf{T} \Hmat_{FF}^{-1} \bfone\right)
 \end{equation}
Where $\det (\Hmat_{FF} )>0$ since $\Hmat_{FF}$ is positive definite and $\det \left(0- \bfone^\mathsf{T} \Hmat_{FF}^{-1}\bfone\right) = - \bfone^\mathsf{T} \Hmat_{FF}^{-1}\bfone  \neq 0$ since $\Hmat_{FF}^{-1}$ is positive definite. Hence, the solution to system~\eqref{eq:KKT-subpb-mat} exists and is unique. It is denoted $(\tilde{\x}^*,\, \lambda^*)$, and will be used in the next step.

\subsection{Keeping primal feasibility}\label{subsec2}

Let $\tilde{\x}$ be the current feasible iterate, 
and $\tilde{\x}^*$ the minimizer obtained by~\eqref{eq:KKT-subpb-mat}. The direction $\bm{d}=\tilde{\x}^*-\tilde{\x}$ satisfies $\bfone^\mathsf{T} \bm{d}=0$ (as the initial and end points both respect the sum-to-$s$ constraint), and $\bm{d}_L=\bfzero$ by construction. Moving along direction $\bm{d}$ with $\x^{(\alpha)} :=\tilde{\x}+\alpha \bm{d} $ decreases the cost function as $\alpha$ increases from 0 to 1, and solutions remain feasible as long as $x^{(\alpha)}_i \geq0$ for all $i$, so the largest feasible step reads:
\begin{equation}\label{eq:alpha_choice}
    \alpha^* = \min_{i\in D^- }\frac{-\tilde{x}_i}{d_i} \text{, where  $D^-:=\left\{ i \in F \ \middle| \ d_i < 0 \right\}$.} 
\end{equation}

Let $j$ denote the argument of the minimum in equation~\eqref{eq:alpha_choice} (should multiple indices allow to reach the minimum, we just keep a randomly chosen one). By definition $\tilde{x}_j=0$, so this step terminates by updating the corresponding set of active constraints:
\begin{equation}
    L =  L \cup \{j\},\quad F = F\setminus \{j\}.
\end{equation}

\subsection{Checking dual feasibility}\label{subsec3}

This step starts with a primal feasible point $\tilde \x^*$ and  $\lambda^*$, solving subproblem~\eqref{eq:KKT-subpb}. Dual feasibility now is required to be checked for variables in $L$, since for $i\in F$ the $\mu_i=0$ by construction. The  vector of Lagrange multipliers is computed with equation~\eqref{eq:KKT-sta-L}:
\begin{equation}
    \bm{\mu}^*_L = \Hmat_{LF} \tilde \x^*_F - \h_L + \lambda^* \bfone
\end{equation}

Then for all $i \in L$ we check the dual feasibility: $\mu^*_i\ge 0$. If this holds, all KKT conditions are satisfied and $\tilde \x^*$ is the global minimizer. Otherwise, we choose the index $j\in L$ with the most negative associated $\mu^*_j$, and it is moved from the active set to the free set:
\begin{equation}
    L= L\setminus\{j\},\quad F= F\cup\{j\}.
\end{equation}

Note that this works because having a negative $\mu^*_j$ means that the gradient along the direction that increases $\tilde x_j$ is negative. Therefore, allowing $\tilde x_j$ to move above zero will decrease the objective while maintaining the sum-to-$s$ equality constraint.

\newpage
\subsection{Flowchart overview}

\hspace{0.7cm}The algorithm described in section~\ref{sec3} --~for which each step was respectively detailed in subsections~\ref{subsec1}, \ref{subsec2}, and~\ref{subsec3}~-- is summarized in figure~\ref{fig:overview} below:

\newcommand{\flowlw}{0.42pt}

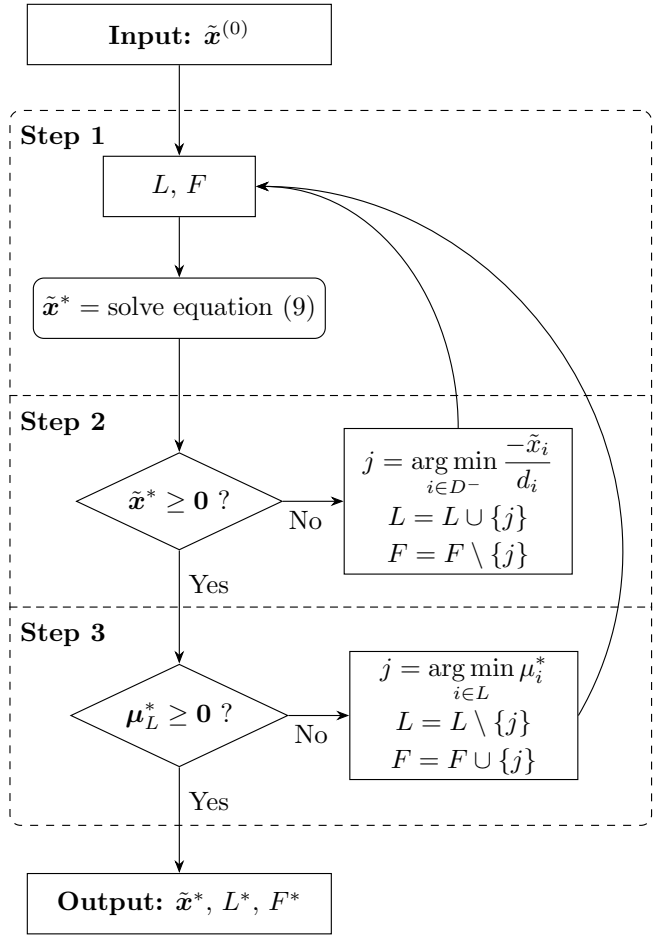
\begin{figure}[!h]
\centering

\scalebox{1}{
\begin{tikzpicture}[
  node distance = 8mm and 24mm,
  >=Stealth,
  every node/.style = {font=\normalsize},
  every path/.style = {line width=\flowlw},
  block/.style      = {draw, line width=\flowlw, rectangle, minimum height=8mm, minimum width=2cm, align=center},
  roundblock/.style = {draw, line width=\flowlw, rectangle, rounded corners, minimum height=8mm, minimum width=4.5, align=center},
  decision/.style   = {draw, line width=\flowlw, diamond, aspect=2.1, align=center, inner ysep=2pt, inner xsep=6pt},
  sideblock/.style  = {draw, line width=\flowlw, rectangle, align=center, inner ysep=2pt, inner xsep=6pt, minimum width=3cm},
  stepmark/.style   = {font=\bfseries, anchor=west, inner sep=1pt},
  dashedframe/.style= {draw, dashed, line width=\flowlw, rounded corners, inner ysep=6mm, inner xsep=3mm, fill opacity=0, text opacity=1}
]

\node[block, minimum width=4cm] (input) {\textbf{Input:} $\tilde \x^{(0)}$};

\node[block, below=12mm of input] (state) {$\displaystyle L,\, F$};

\node[roundblock, below=of state] (solve) {$\displaystyle \tilde \x^* = \text{solve equation~\eqref{eq:KKT-subpb-mat}}$};

\node[decision, below=15 mm of solve] (check1) {$\displaystyle \tilde \x^* \geq \bfzero \text{ ?}$};

\node[sideblock, right=8mm of check1] (blk1) {$\displaystyle j = \argmin_{i\in D^- }\frac{-\tilde{x}_i}{d_i}$\\[2pt]
$\displaystyle L =  L \cup \{j\}$\\[2pt]
$F = F\setminus \{j\}$};

\node[decision, below=15 mm of check1] (check2) {$\displaystyle \bm{\mu}^*_L \geq \bfzero \text{ ?}$};

\node[sideblock, right=8mm of check2] (blk2) {$\displaystyle j = \argmin_{i\in L}\mu^*_i$\\[2pt]
$\displaystyle L =  L \setminus \{ j\}$\\[2pt]
$F = F\cup \{ j\}$};

\node[block, minimum width=4cm, below=14mm of check2] (output) {\textbf{Output:} $\tilde \x^*,\, L^*,\, F^*$};

\draw[->] (input) -- (state);
\draw[->] (state) -- (solve);
\draw[->] (solve) -- (check1);
\draw[->] (check1) -- node[right, yshift=0.31cm]{Yes} (check2);
\draw[->] (check2) -- node[right, yshift=0.25cm]{Yes} (output);

\draw[->] (check1.east) -- node[below, xshift=-0.1cm]{No} (blk1.west);
\draw[->] (check2.east) -- node[below, , xshift=-0.1cm]{No} (blk2.west);

\draw[->] (blk1.north) to[out=90, in=0, looseness=1] (state.east);
\draw[->] (blk2.east) to[out=60, in=0, looseness=1.1] (state.east);

\coordinate (padR) at ([xshift=7mm] $(blk1.east)!0.5!(blk2.east)$);
\node[dashedframe, fit=(state) (solve) (check1) (blk1) (check2) (blk2) (padR)] (frame) {};

\path let \p1 = (frame.west), \p2 = (frame.east) in
  coordinate (leftedge)  at (\x1,0)
  coordinate (rightedge) at (\x2,0);
\coordinate (mid34) at ($(solve.south)!0.5!(check1.north)$);
\coordinate (mid56) at ($(check1.south)!0.5!(check2.north)$);
\draw[densely dashed] (leftedge |- mid34) -- (rightedge |- mid34);
\draw[densely dashed] (leftedge |- mid56) -- (rightedge |- mid56);

\node[stepmark] at ([yshift=-0.37cm, xshift=0.1cm] frame.north west) {Step 1};
\node[stepmark] at ([yshift=-0.37cm, xshift=0.1cm] leftedge |- mid34) {Step 2};
\node[stepmark] at ([yshift=-0.37cm, xshift=0.1cm] leftedge |- mid56) {Step 3};

\end{tikzpicture}
}\caption{Flowchart overview of the active-set algorithm}\label{fig:overview}
\end{figure}

\section{References}
\renewcommand{\refname}{}
\bibliographystyle{myieeetr}
\vspace{-2em}
\bibliography{References}

\end{document}